# Total electron content and electron density profile observations during geomagnetic storms using COSMIC satellite data

Niraj Bhattarai[1], Narayan P Chapagain[1], Binod Adhikari[2]

1.Patan Multiple Campus, PatanDhoka, Lalitpur, Tribhuvan University, Nepal
2.Amrit Science Campus, Kathmandu, Tribhuvan University, Nepal

## ABSTRACT

Total Electron Content (TEC) and electron density are the basic parameters, which determine the major properties of the Ionosphere. Detail study of the ionospheric TEC and electron density variations has been carried out during geomagnetic storms, with longitude and latitude, for four different locations: (24°W-14°W, 25°S-10°S); (53°W- 46°W, 04°N-14°N); (161°E-165°E, 42°S-34°S), and (135°W-120°W, 39°S-35°S) using the COSMIC satellite data. In order to find the background conditions of the ionosphere, the solar wind parameters such as north-south component of inter planetary magnetic field (Bz), plasma velocity (Vsw), AE, Dst and Kp indices, have also been correlated with the TEC and electron density. The results illustrates that the observed TEC and electron density profile significantly vary with longitudes and latitudes as well. This study illustrates that the values of TEC and the vertical electron density profile are influenced by the solar wind parameters associated with solar activities. The peak value of electron density and TEC increase as the geomagnetic storms becomes stronger. Similarly, the electron density profile vary with altitudes which peaks around the altitude range of about 180-280 km, depending on the strength of geomagnetic storms. The results clearly show that the peak electron density shifted to higher altitude (from about 180 km to 300 km) as the geomagnetic disturbances becomes stronger.

## 1. INTRODUCTION

In the outer earth's environment, solar radiation strikes the atmosphere with high density. The intense level of radiation is spread over a spectrum ranging from radio frequency through IR (Infrared Radiation and X-ray [1]. The intense X-ray and UV ionized region of the atmosphere is plasma containing the charged particles; electrons and ions. The ionizing action of the sun's radiation on the earth's upper atmosphere produces free electrons. Above about 60 km, the number of these free electrons is sufficient to affect the propagation of electromagnetic waves [1].

The number of free electrons along the ray path of one-meter-squared cross section that extends all the way up from the ground through the ionosphere is referred as total electron content [TEC] [2,3],or the electron columnar number density. Its unit TECU, 1 TECU equals to $10^{16}$ electrons per square meter [2]. A method of the calculation of the TEC using the dual GPS satellite system is found to be discussed in a paper by *Shim* [2]. The constituents and chemical composition of the ionosphere as a whole is not same. It varies with altitude and temperature. The main composition of the D region is $N_2$, $O_2$, Ar, $CO_2$, He, besides this highly variable quantity of $O_3$ and $H_2O$ are also present, [4, 5, 6, 7, 8]. And dust particles sustaining in the atmosphere along with some traces of Iron and Sodium received from the meteor debris. There are some other negative ions like$O2^-$, electrons, etc. In the E and F region mainly the molecular ions like$NO^+$, $O2^+$,$O^+$and $N2^+$are dominating. In the upper F region ions like$O^+$, $H^+$and $He^+$are present [8]. This paper pointed out that the dynamics of the ionosphere and the variation in the plasma density was due to prolonged exposure to the low-level extreme ultra violet (EUV) radiation. Also the causes of ionization of neutrals and molecules of the different regions of the ionosphere are proved to be governed by solar UV rays, X-Rays, Lyman alpha radiation and galactic cosmic rays [9].Furthermore, the response of low-latitude ionosphere to the geomagnetic storm of 24 August 2005 was discussed by *Sharma et al.* [3]. They reported the ionospheric influence of the different geomagnetic disturbances and solar wind parameter observed with the different events of TEC values.

During a geomagnetic disturbance, there is an energy input inside the magnetosphere and ionosphere, which changes ionospheric parameters, such as composition, temperature and circulation. But during quiet periods, measurements on the ground do not present significant disturbances. Geomagnetic storms are initiated due to the activities on the Sun, mainly by the flares and/or the coronal mass ejection, when the solar wind velocity, temperature and density vary drastically, accompanied by significant changes in the north-south component of the interplanetary magnetic field (IMF,Bz) [3]. This paper have studied the response of low-latitude ionosphere to the geomagnetic storm of 24 August 2005 in terms of variations in TEC. Also the latitudinal variations in TEC from anomaly crest region in the northern hemisphere down to near equatorial stations that lie within 75°E ± 3°E (Local Time, LT = UT + 0500, at 75°E) and longitudinal variation in TEC for the GPS satellite data has been studied. In a paper [10] a comparative study of TEC was carried out using different satellite data. The paper also reviewed the progress in the development of tomographic inversion techniques using total electron content measurements to image the ionosphere as an aid to various radio systems applications.

In this paper, we have studied the TEC and electron density profile of the ionosphere for four different geographical locations using COSMIC satellite data by choosing cases of different geomagnetic storms. For this we have selected four different solar and geomagnetic storms time in the years 2011, 2012 and 2014 including one geomagnetic quiet time event. Planetary-scale magnetic activity is measured by the Kp index and south- north component of interplanetary magnetic field, Bz,[11, 12], which is the basis of the event selection. In those events, we have investigated the storm effect in the ionosphere by observing the variation of TEC with latitude and longitude and electron density profile with respect to altitude, longitude and latitude for each event. At last a collective comparison of the electron density profile for all the events taken together has been analyzed.

## 2. DATA AND MEASUREMENTS

We have used Internet based supply of data provided by Operating Mission as Nodes on the Internet web system (OMNI) and Constellation Observing System for Meteorology, Ionosphere, and Climate (COSMIC) satellite system. From OMNI system we have used the data for the events selection. And for the selected date time and geographical locations, we have taken data supply of the TEC and electron density profile from the COSMIC Satellite system. These data are processed and provided to the thousands of the users by COSMIC data analysis and archive center (CDAAC) of the university corporation for atmospheric research (UCAR). The date and the time of the data taken is tabulated in Table-1.

**Table 1** Selection of the events

| Events | Date | Time (UT) |
|---|---|---|
| Event-1 | 16 Aug 2014 | 1941 |
| Event-2 | 27 Aug 2014 | 1122 |
| Event-3 | 14 Nov 2012 | 0302 |
| Event-4 | 26 Sep 2011 | 1801 |

### 2.1. OMNI and COSMIC System

OMNI project is an internet based data explorer system developed by the National Aeronautics Space Administration (NASA). From OMNI system we have used data observation of horizontal component of the magnetic field (Bz), plasma velocity (Vsw) and indices- Kp, AE and Dst. The detail about these parameters can be found in paper of [13, 14, 15, 16]. These data are made freely accessible from the web page http://omniweb.gsfc.nasa.gov/ow.html.

In 15 April 2006, 0140 UTC, six microsatellites were launched into a circular, 72° inclination orbit at an altitude of 512 km from Vandenberg Air Force Base, California [17]. The mission is a collaborative project of the National Space Organization (NSPO) in Taiwan and the UCAR in the United States. The mission is called COSMIC in the United States and the same mission is named as Formosa Satellite Mission 3 (FORMOSAT-3) in Taiwan. After launch the six satellites were orbiting very close to each other at the initial altitude of 512 km. During the first 17 months following their launch the satellites were gradually dispersed into their final orbits at ~800 km, with a separation angle between neighboring orbital planes of 30° longitude [18]. The primary payload of each COSMIC satellite is a GPS radio-occultation receiver developed by the NASA's Jet Propulsion Laboratory (JPL). By measuring the phase delay of radio waves from GPS satellites as they are occulted by the Earth's atmosphere, accurate and precise vertical profiles of the bending angles of radio wave trajectories are obtained in the ionosphere, stratosphere, and troposphere. From the bending angles, profiles of atmospheric refractivity (*N*) is obtained [18]. The refractivity *N* is a function of temperature (*T*), pressure (*p*), water vapor pressure (*e*), and electron density ($n_e$). The relation for electron density is given by

$$n_e = \left[1.92 \times 10^{-6} \frac{p}{T} + 9.2 \times 10^{-3} \frac{e}{T^2} - 2.5 \times 10^{-8} N\right] f^2$$

Where, *f* is the frequency of the GPS carrier signal (Hz). The detail method of the calculation of the TEC for the COSMIC satellite system is described by *Yue et al.*[19]. The current orbital configuration of COSMIC system is giving global coverage of approximately 2,000 observation soundings per day, distributed nearly uniformly in local solar time. The observations data are obtained from the website of the Taiwan Analysis Center for COSMIC (TACC) in Taiwan at http://tacc.cwb.gov.tw, and also available through the UCAR Web site at www.cosmic.ucar.edu/. In this work we have used the essential data from the above mentioned site. The data has been analyzed using MATLAB program.

### 3. RESULTS AND DISCUSSIONS

In this section we have described the observations and result of four different events. The event 1 is selected for quiet day, while the other 3 events are selected from geomagnetic storms days. We have plotted TEC and electron density profile with longitude and latitude and electron density profile with altitude for each event and finally discussed their comparison studies.

### 3.1. Observed Solar Wind Parameters

Figure-1 represents the OMNI datasets during the Event-1 from 15-17 August 2014. According to *Gonzalez et al.* [16], geomagnetic storms can be classified as: weak (−50 <Dst≤ −30nT), moderate (−100 <Dst≤ −50nT), intense (−250 <Dst≤ −100nT), and very intense (Dst≤ −250nT). In this figure, the value of Bz lies in between -4 nT to 5 nT, plasma speed is around 300km/s, Kp value do not exceed 3, Dst index lies near zero with small negative values and the AE index is below 100 nT. Thus the all solar wind parameters and the geomagnetic indices have low values, with characterizing Event-1 a quiescent condition or quiet event. The values of solar wind parameters are observed to vary throughout the time interval between 15 to 17 August but our work is focused only in a point of the time interval represented by double arrow in the AE panel.

Figure-2 represents the OMNI datasets during the Event-2 from 26-28 Aug 2014. In this figure, the geomagnetic indices present moderate values. The value of Bz lies in between -10 nT to 4 nT, plasma speed shows a consistent value of about 300km/s. Kp value do not exceed 5. The Dst index shows a moderate value of -75 nT at its least and the AE index lies in between zero to 1000 nT at most. Thus the solar wind parameters and interplanetary magnetic field show moderate values, for Event-2. In the first panel of the figure we can observe that the value of Bz is about -10 nT for some hours. This range is marked by the double arrow drawn in the AE panel. In this event, our work is focused to observe the variation of TEC and electron density in the time interval represented by double arrow in the AE panel.

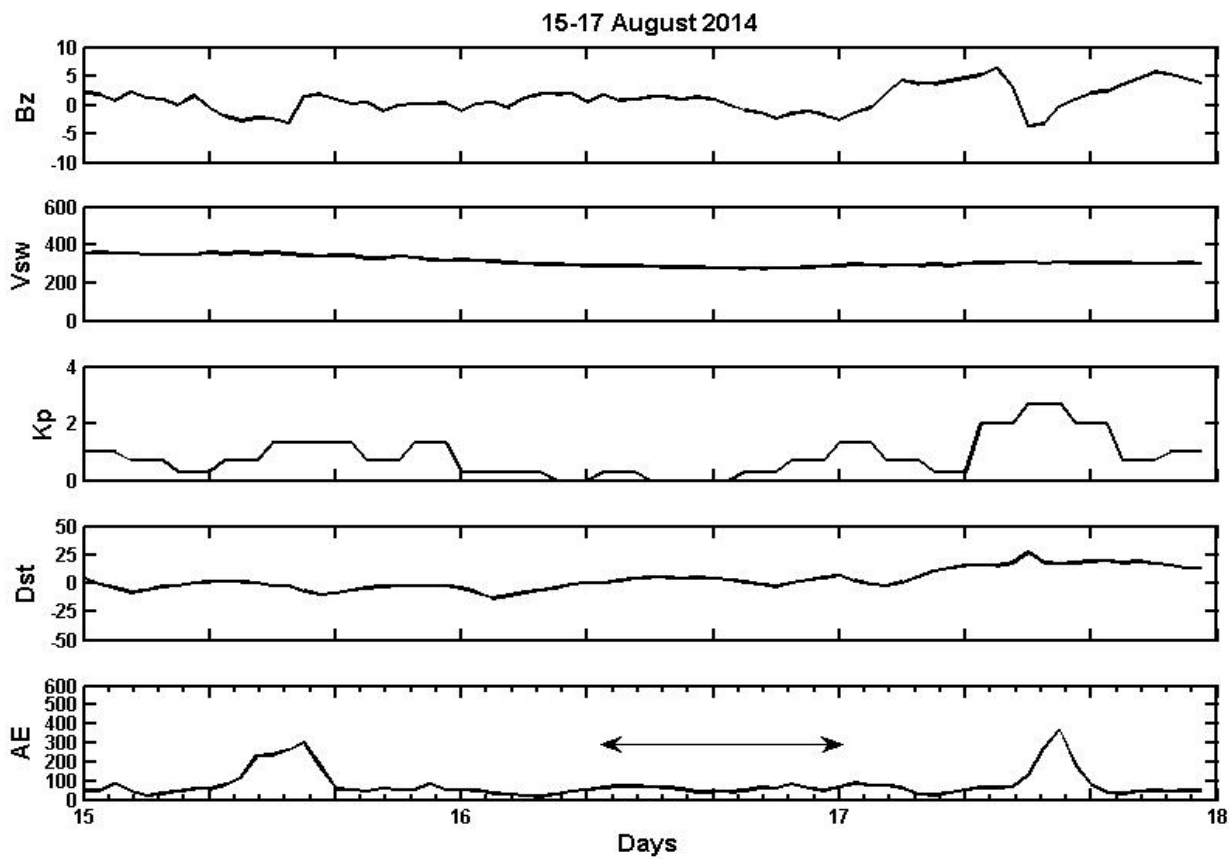

**Figure 1** From top to bottom, the panels shows the variations of the south-north component of magnetic field Bz (nT) in GSM coordinate system, solar wind plasma speed, Vsw, (km/s), Kp, Dst (nT) and AE (nT) indices with time (days) respectively. The arrow in the AE panel marks the selected time interval for Event-1, 15-17 August 2014.

Figure-3 represents the OMNI datasets during the Event-3 from 13-15 November 2012. In this figure, the value of Bz lies in between -17nT to 17nT. The maximum plasma speed is observed to about 450km/s. Similarly Kp index value lies in between zero and 5. The Dst index starting from zero falls to -110 nT to its minimum value and at last panel we observed the AE index lies in between 0nT and 1000nT. Bz reaches to -17nT for some hours as indicated by the double arrow in the AE panel. We study the variation of the TEC and electron density of a point at this particular time interval. Similarly, Figure-4 represents the OMNI datasets during the Event-4 from 25-27 September 2011. In this figure, the south-north component of interplanetary magnetic field (Bz) lies in between -24 nT to 10nT. Similarly other solar wind parameters like plasma speed shows a range of 300 to 700 km/s. Kp index has value in between zero to 7. The Dst index reaches a deepest value of -120nT starting from zero and the bottom panel shows the variation of AE index from zero to a peak value of 1800 nT. Bz is observed to approach -24nT for only a short interval of time, which is indicted by the arrow in the AE panel. Our work is focused to observe the variation of TEC and electron density with longitude and latitude and also electron density variation with altitude at a point during this interval.

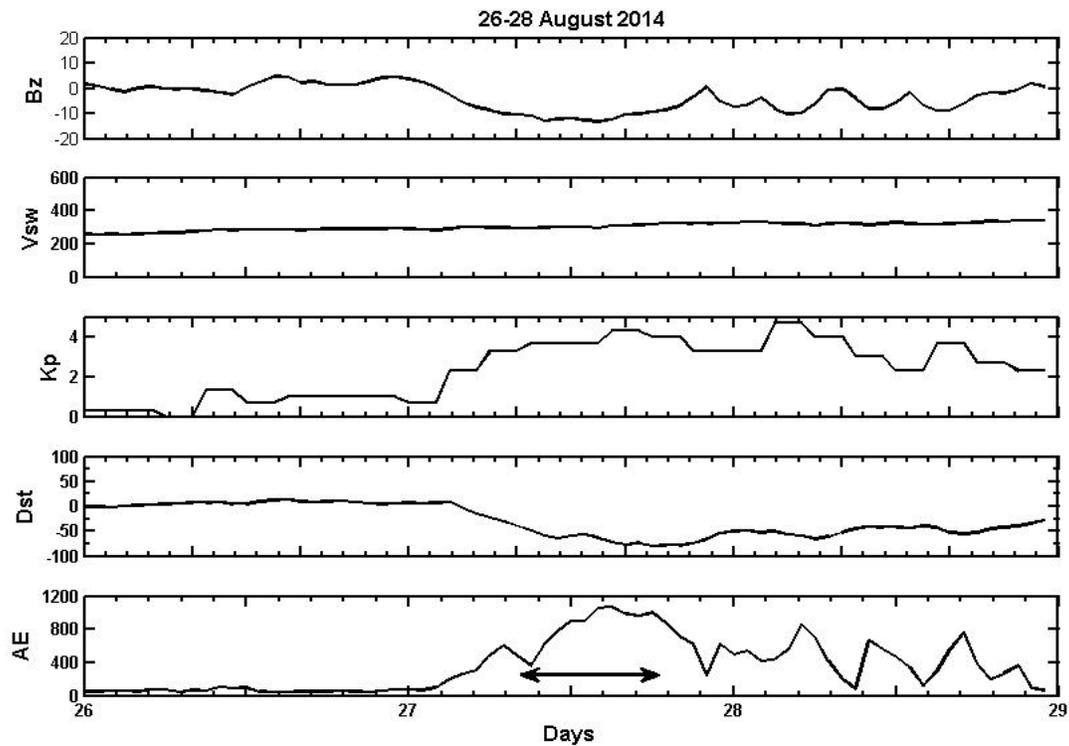

**Figure 2** From top to bottom, the panels shows the variations of the south-north component of magnetic field Bz (nT) in GSM coordinate system, solar wind plasma speed, $V_{sw}$, (km/s), Kp, Dst (nT) and AE (nT) indices with time (days) respectively. The arrow in the AE panel marks the selected time interval for Event-2, 26-28 August 2014.

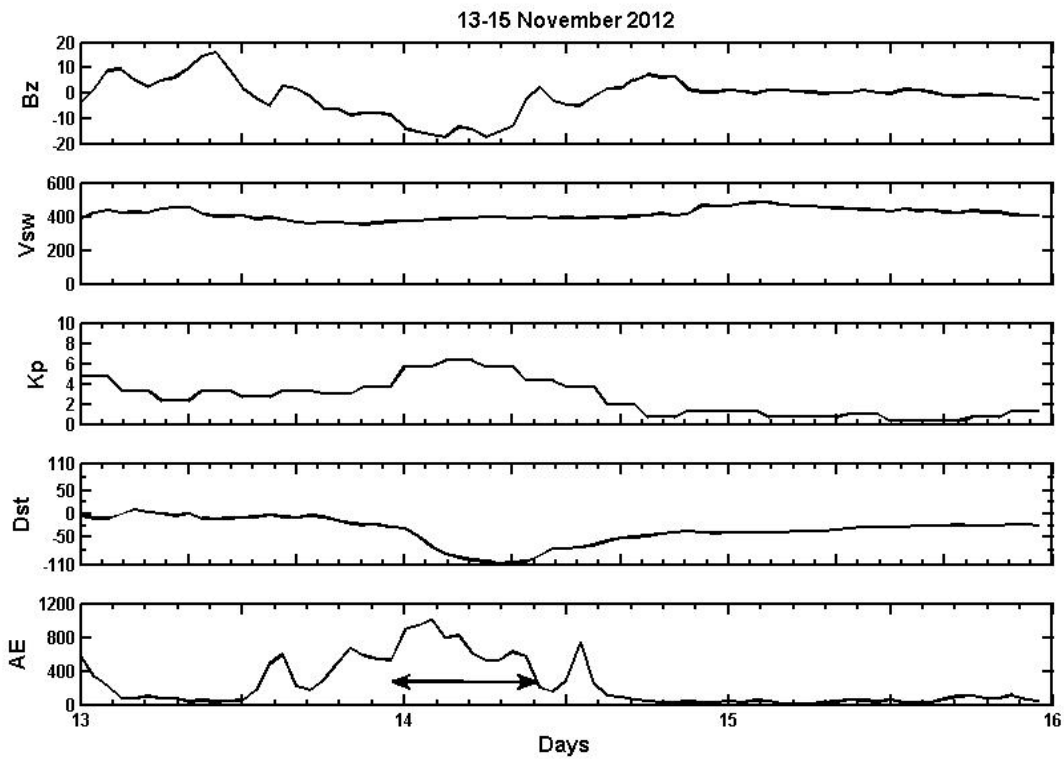

**Figure 3** From top to bottom, the panels shows the variations of the south-north component of magnetic field Bz (nT) in GSM coordinate system, solar wind plasma speed, Vsw, (km/s), Kp, Dst (nT) and AE (nT) indices with time (days) respectively. The arrow in the AE panel marks the selected time interval for Event-3, 13-15 November 2012.

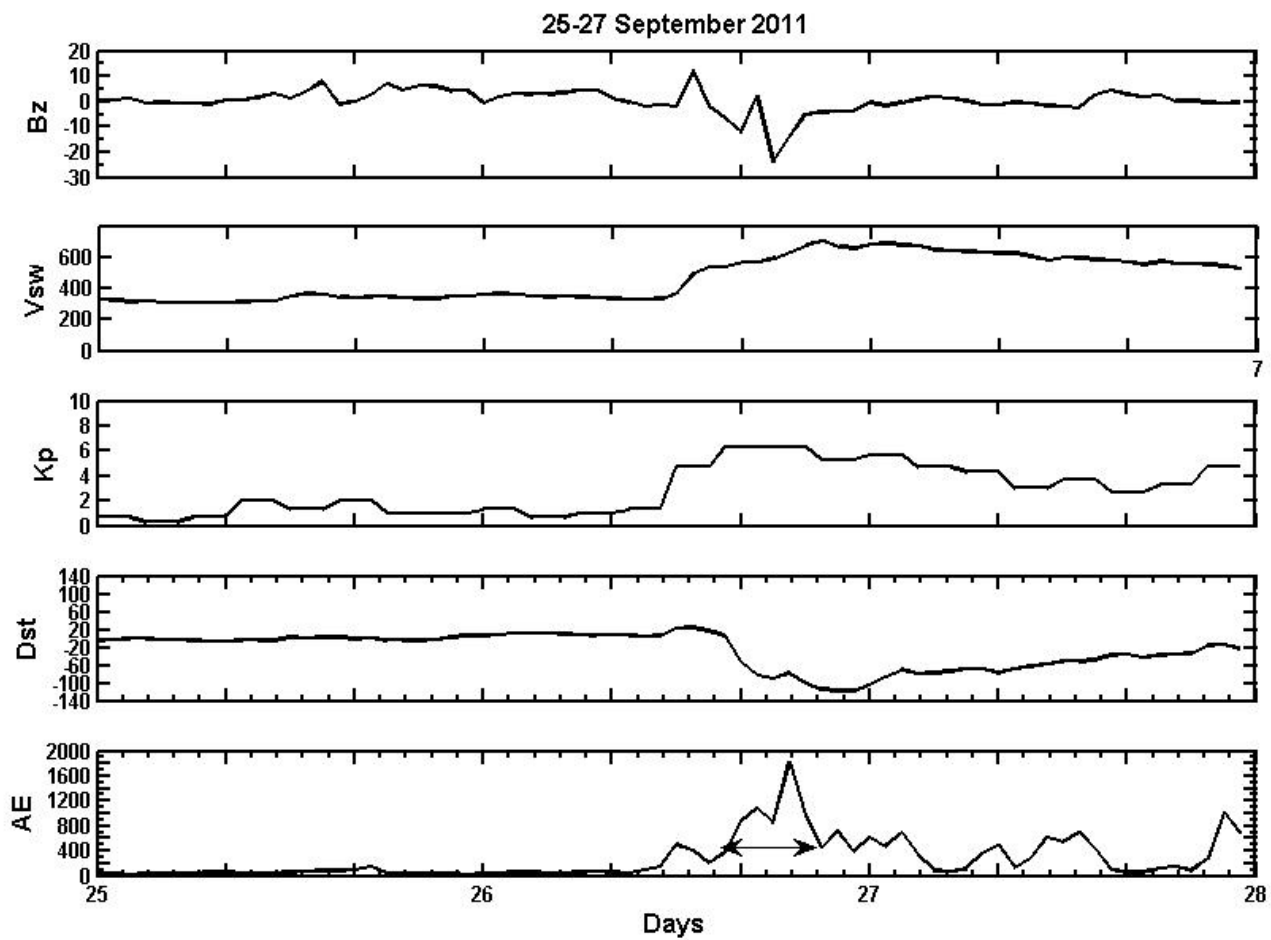

**Figure 4** From top to bottom, the panels shows the variations of the south-north component of magnetic field Bz (nT) in GSM coordinate system, solar wind plasma speed, $V_{sw}$, (km/s), Kp, Dst (nT) and AE (nT) indices with time (days) respectively. The arrow in the AE panel marks the selected time interval for Event-4, 25-27 September 2011.

### 3.2. Observed TEC and Electron Density Variations

Plots (a) and (b) in Figure-5 show the variations of TEC and electron density with latitudes (range 25°S to 10°S), respectively. With decrease in latitudes (from latitude 25°S to 10°S), both the TEC and electron density gradually increased. TEC increases from nearly zero to a peak value of 120 TECU and electron density exponentially increases from a low value of about $0.1 \times 10^5$ to $6 \times 10^5 elcm^{-3}$. Both TEC and electron density increase towards the high-latitude regions and peaks are obtained at latitude about 12°S. Plots (c) and (d) in Figure-5 show the variations of TEC and electron density with longitude (range 24°W to 14°W), respectively. As we go from longitude 24°W to 14°W, we see the TEC and electron density value gradually increasing, where TEC increases from nearly zero to a peak value of about 120 TECU and electron density reaches a peak value $6 \times 10^5 elcm^{-3}$. Both TEC and electron density reachits peak value at 16°W of prime meridian.

Plots (a) and (b) in Figure-6 show the variations of TEC and electron density with latitude (range 04°N to 14°N) respectively. As we go from latitude 04°N, to 14°N, we see the TEC and electron density gradually increasing, where TEC increases from near zero to a peak value of 200 TECU and electron density reach a peak value of $7.8 \times 10^5 elcm^{-3}$ and then both decrease on moving further right. Both TEC and electron density reaches its peak value at 12°N. Plots (c) and (d) in Figure-6 show the variations of TEC and electron density with longitude (range 53°W to 46°W) respectively. We see peak values of TEC and electron density at around longitude 51°W of prime meridian, and both of them gradually decreases as we move either side of that location. The peak TEC is observed to be 200 TECU and density is observed to be $7.8 \times 10^5 elcm^{-3}$.

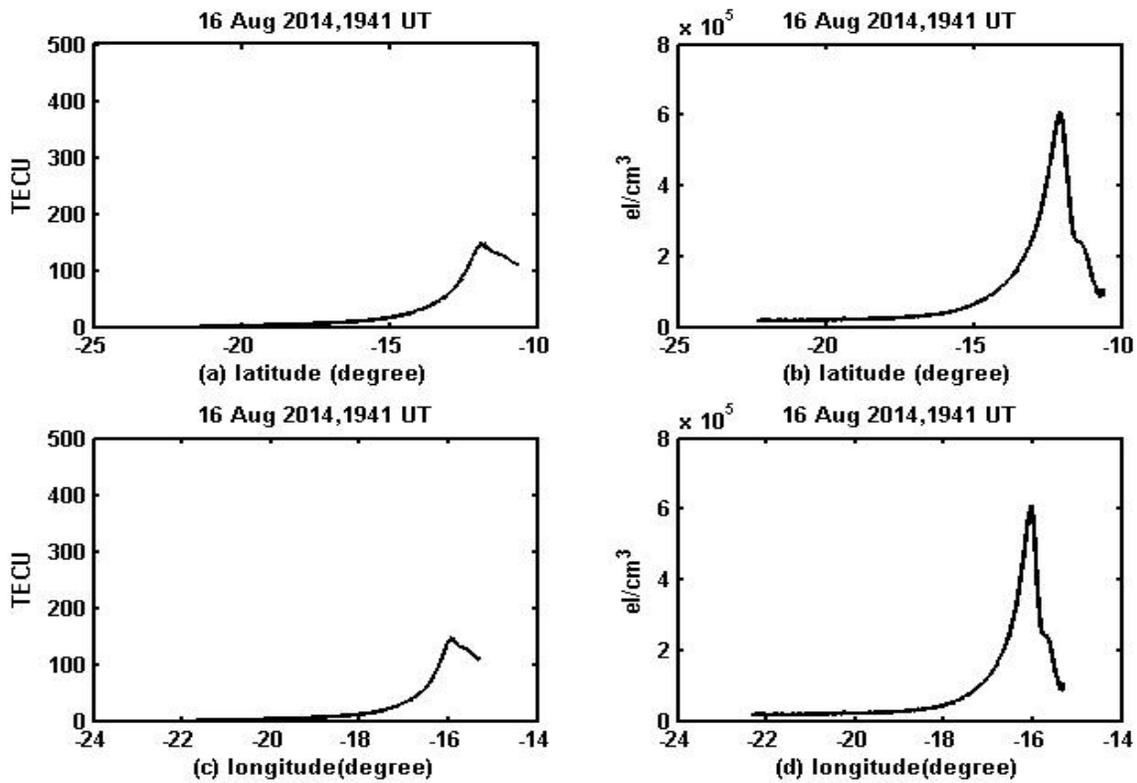

**Figure 5** The variation of, (a) TEC (TECU) with latitude (degree), (b) electron density (elcm$^{-3}$) with latitude (degree), (c) TEC (TECU) with longitude (degree) and (d) electron density (elcm$^{-3}$) with longitude (degree) from Event-1 on 16 August 2014, 1941 UT.

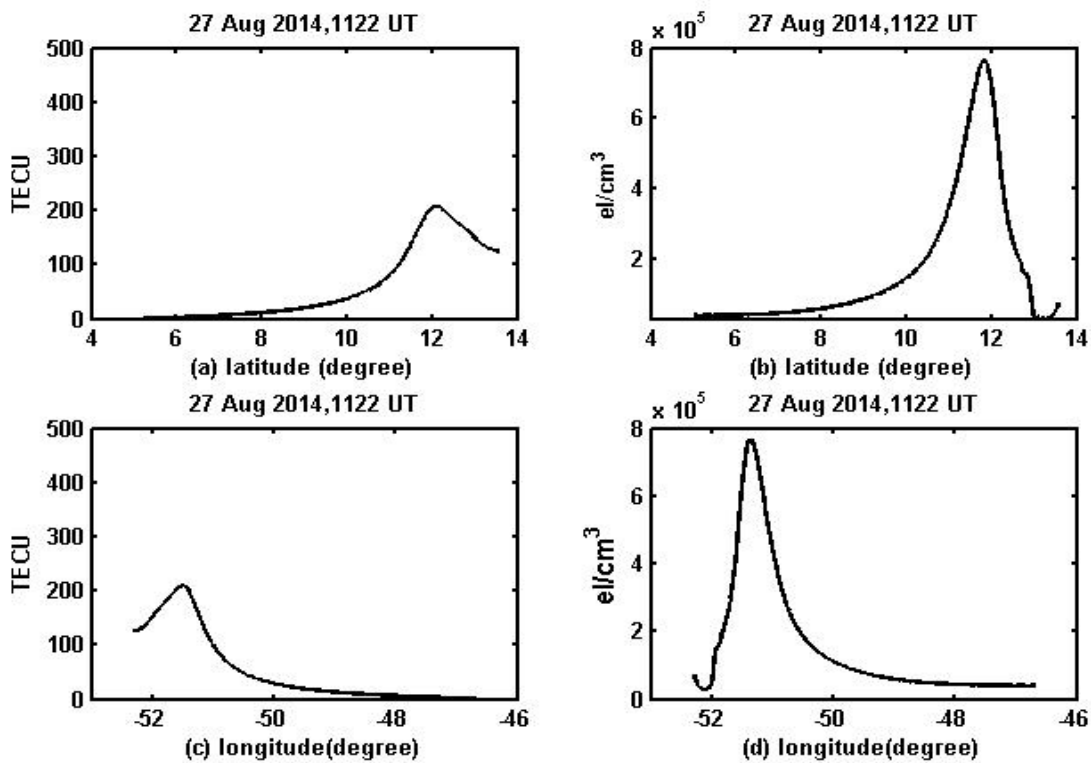

**Figure 6** The variation of, (a) TEC (TECU) with latitude (degree), (b) electron density (elcm$^{-3}$) with latitude (degree), (c) TEC (TECU) with longitude (degree) and (d) electron density (elcm$^{-3}$) with longitude (degree), Event-2, 27 August 2014, 1122 UT.

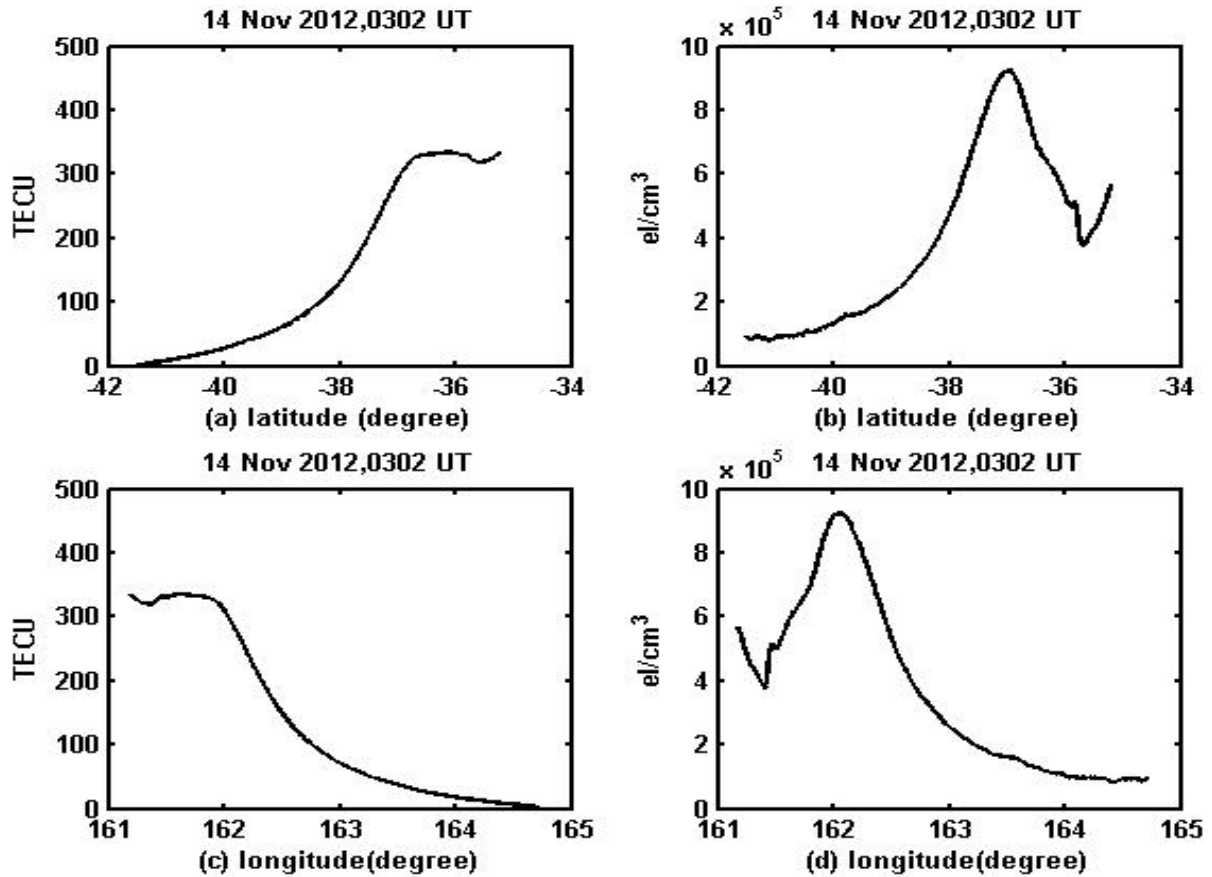

**Figure 7** The variation of, (a) TEC (TECU) with latitude (degree), (b) electron density (elcm$^{-3}$) with latitude (degree), (c) TEC (TECU) with longitude (degree) and (d) electron density (elcm$^{-3}$) with longitude (degree), Event-3, 14 November 2012, 0302 UT.

Plots (a) and (b) in Figure-7 show the variations of TEC and electron density with latitude (range 42°S to 34°S) respectively. As we go from latitude 42°S to 34°S we see the TEC and electron density value gradually increasing, where TEC increases from near zero value to a peak value of 320 TECU and electron density reach a peak value of $9\times10^5$ elcm$^{-3}$ and then density decrease on further moving right whereas TEC remains almost constant for a while and shows a sudden decrease. Both TEC and density reach its peak value at 37°S. Plots (c) and (d) in Figure-7 show the variations of TEC and electron density with longitude (range 161°E to 165°E) respectively. We see peak values of TEC and electron density at around longitude 162°E. TEC initially at its peak value at 161°E, decrease gradually on moving higher longitude, whereas electron density decreases as we move either side of 162°E. The peak value TEC is 320 TECU and density is $9\times10^5$ elcm$^{-3}$. Plots (a) and (b) in Figure-8 show the variations of TEC and electron density with latitude (range 39°S to 35°S) respectively. As we go from 39°S to 35°S latitude, we see the TEC and electron density value increasing, where TEC increases from its initially considerable value to a peak of 380 TECU. Similarly electron density increase and reach a peak value of $11.5\times10^5$ elcm$^{-3}$. Beyond these peak point, TEC as well as density curve shows gradual decrease in values on further moving right. Both TEC and electron density reaches its peak value at 38°S latitude.

Plots (c) and (d) in Figure-8 show the variations of TEC and electron density with longitude (range 135°W to 120°W) respectively. With the decrease in longitude from 135°W, TEC and electron density slowly increases to a peak value and then starts decreasing on going further above. The peak values of TEC and electron density are observed to be 380 TECU and $11.5\times10^5$ elcm$^{-3}$ at around longitude 133° W of prime meridian respectively. Table 2 summarizes results of the events 1-4 for values Bz, Kp, electron density, and TECU obtained from Figure-5 through Figure-8. It shows that the peak value of electron density for Event-1 to Event-4 are $6\times10^5$ elcm$^{-3}$, $7.8\times10^5$ elcm$^{-3}$, $9\times10^5$ el/cm$^3$ and $11.5\times10^5$ el/cm$^3$ and peak value of TEC are 120, 200, 320 and 380 TECU respectively. These peak values are observed at (latitude, longitude) = (12°S, 16°W), (12°N, 51°W), (37°S, 162°E), and (38°S, 133° W) of Event-1 to Event-4 respectively. The results show that with increase of the strength of the geomagnetic storms from event 1 to event 4, the values of electron density and TECU also increases. The nature of the variations of electron density and TECU we observed are in good agreement with the previous studies [10, 20, 3].

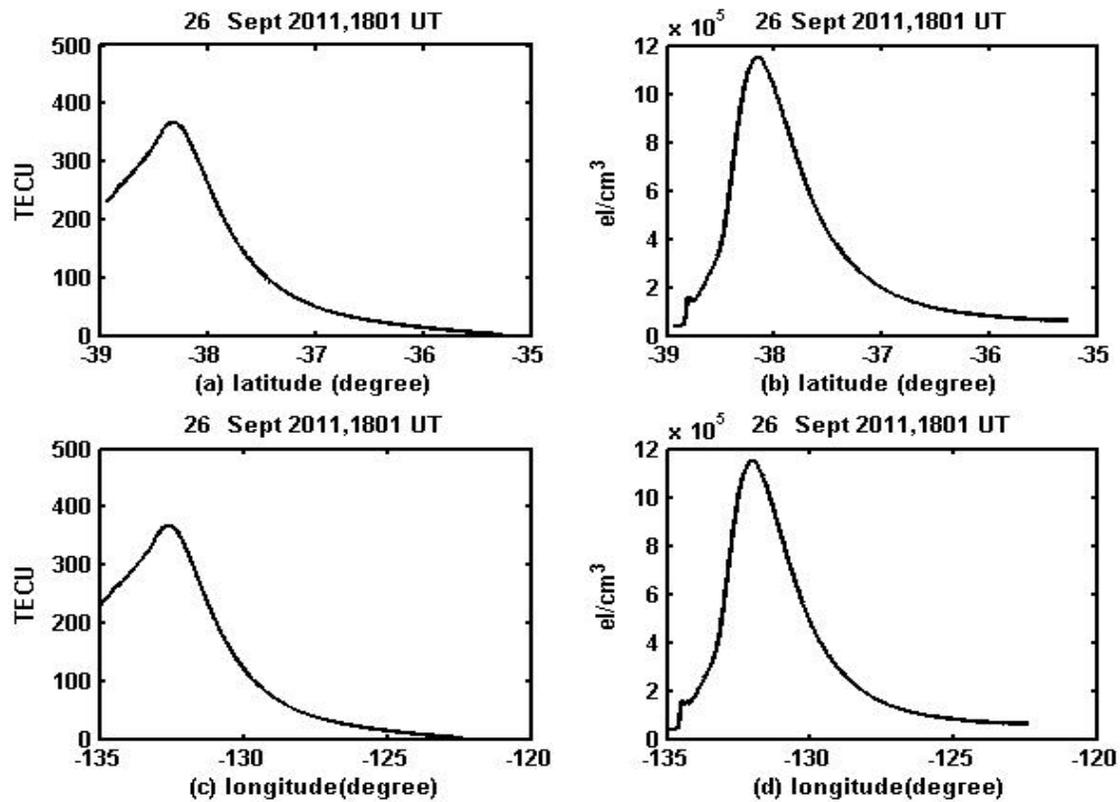

**Figure 8** The variation of, (a) TEC (TECU) with latitude (degree), (b) electron density (elcm$^{-3}$) with latitude (degree), (c) TEC (TECU) with longitude (degree) and (d) electron density (elcm$^{-3}$) with longitude (degree), Event-4, 26 September 2011, 1801 UT.

**Table 2** Summary of events 1-4 for Bz, Kp, electron density, and TECU

| Events | Date/Time (UT) | Geo Position | Bz (nT) | Kp | Electron Density(elcm$^{-3}$) | TECU |
|---|---|---|---|---|---|---|
| 1 | 16 August 2014 1941 UT | Lon (24˚S-14˚S)Lat (25˚W-10˚W) | -2 | 1 | $6 \times 10^5$ | 120 |
| 2 | 27 August 2014 1122 UT | Lon (53˚W-46˚W) Lat (4˚N-14˚N) | -10 | 4 | $7.8 \times 10^5$ | 200 |
| 3 | 14 Nov 2012 0302 UT | Lon (161˚E-165˚E) Lat (42˚S - 34˚S) | -17 | 6 | $9 \times 10^5$ | 320 |
| 4 | 26 Sep 20111801 UT | Lon(135˚W-120˚W) Lat (39˚S -35˚S) | -24 | 7 | $11.5 \times 10^5$ | 380 |

### 3.3. Electron Density Variations with Altitude

Figure-9 shows the variations of electron density with the altitude for the four different events associated with different geomagnetic storms. The plots with blue, green, black and red curves represents for Event- 1, 2, 3 and 4 respectively. Blue curve reflects with the increase in altitude, electron density increases and reaches a maximum of $6\times10^5$ elcm$^{-3}$ at a height of ~180km and then gradually decreases above that height. Green curve shows with the increase in altitude of the ionosphere, electron density also increases and reach a maximum of $7.8\times10^5$ elcm$^{-3}$ at a height of 240 km and then again gradually decreases above that height showing comparatively less electrons present above 600 km. Black curve initially shows some fluctuation of electron density that reaches a peak value of about $9 \times 10^5$ el/cm$^3$ at an altitude 280km and its value drops exponentially on further increasing altitude. The red curve shows with the increase in altitude of the ionosphere, electron density profile also increases and reaches a maximum of $11.5 \times 10^5$ el/cm$^3$ at a height of 280km and then again gradually decreases above that height showing very less electron density profile above 800km.

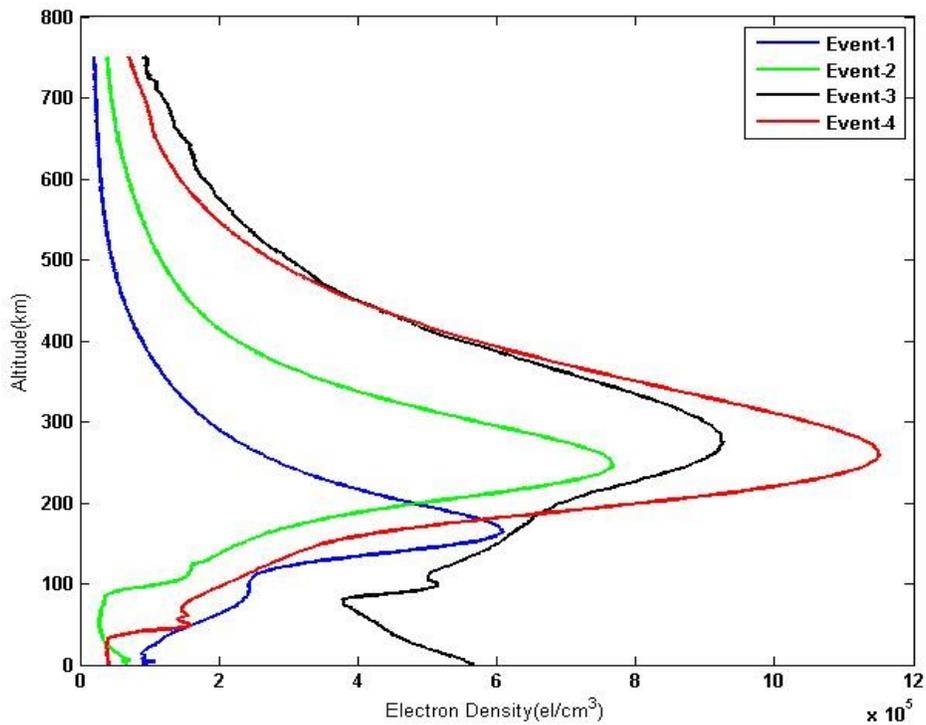

**Figure 9** Comparison of electron density variations with altitude (km), from the four different events

The nature of the curve of electron density profile versus altitude is similar for all the events [Figure-9], but the magnitude of the electron density is increased from Event-1 through 4.These results are in agreement with previous results [10, 20, 3] and the nature of the curve we observed is similar to that reported by *Hajj and Romans* [21], and *Stolle et al.* [22]. In the works of *Hajj and Romans* [21], electron density profile is taken from the GPS/MET (Global Positioning System/ Meterology) and CHAMP (Challenging Minisatellite Payload) radio occultation data. During geomagnetic disturbance there is an energy input inside the magnetosphere and ionosphere, it is predicted that a quiet state of ionosphere starts to be unsettled which change ionospheric parameters, such as composition, temperature and circulation. But during quiet periods, measurements on the ground do not present significant disturbances [11, 23, 24]. Mainly the energy input comes in the form of cosmic rays from the outer space, energetic radiations like UV rays, X-rays and infrared radiations during high solar activities, solar flares, solar storms and many other geomagnetic activities. The magnitude of the radiations received is dependent on the day and night as well as the period of solar cycle. We concluded that when there is high energy input, perturbation of the ionosphere increase and the molecules and atoms receiving sufficient energy ionizes the gases, with thus increasing the number of free electrons. This consequently aids to increase the electron density as well as the TEC.

## 4. CONCLUSIONS

We have studied the variations of electron density profile and TEC with longitude and latitude for the events of different geomagnetic storms using data from ONMI and COSMIC satellite system. For this study, we have used solar wind parameters such as Kp and Dst values, AE index, plasma drift velocity, and interplanetary magnetic field. We selected four different events geomagnetic quiet to strong disturbances from four different locations: (24°W-14°W, 25°S-10°S); (53°W- 46°W, 04°N-14°N); (161°E-165°E, 42°S-34°S), and (135°W- 120°W, 39°S-35°S). The results show that TEC and electron density significantly vary with longitude and latitude. The observed electron density and the TEC follow the similar trend of variations for all longitudes and latitudes, in each event. The peak values of electron density for Event-1 through Event-4 are as: $6\times10^5$ el/cm³, $7.8\times10^5$ el/cm³, $9\times10^5$ el/cm³ and $11.5\times10^5$ el/cm³, respectively. We observed the peak value of electron density and TEC increase on moving from Event-1 through Event-4 as the geomagnetic storms becomes stronger. Similarly, the electron density profile varies with altitudes which peak around the altitude range of about 200-280 km, depending on the strength of geomagnetic storms. The results clearly show that the peak electron density shifted to higher altitude as the geomagnetic disturbances increases. As part of future investigation, we are planning to study the long-term data set of TEC and electron density profile from COSMIC satellite measurements and perform the comparative study from other measurements both from ground- and space-based observations from different regions.

## ACKNOWLEDGEMENTS

This work was supported by Department of Physics, Patan Multiple Campus. And special thanks to the office of B. P. Koirala Memorial Planetarium, Observatory and Science Museum Development Board (BPKMPOASMDB), Kirtipur Kathmandu for providing scholarships to NirajBhattarai to complete his dissertation work.